\documentclass[graybox]{svmult}

\usepackage{amsfonts,amsmath,amssymb}
\usepackage{epstopdf}
\usepackage{mathptmx}       
\usepackage{helvet}         
\usepackage{courier}        
\usepackage{type1cm}        
\usepackage{makeidx}         
\usepackage{graphicx}        
\usepackage{multicol}        
\usepackage[bottom]{footmisc}

\makeindex             

\begin{document}

\title*{Phase transitions in general gravity theories}
\author{Xi\'an O. Camanho}
\institute{Xi\'an O. Camanho \at Department of Particle Physics and IGFAE, University of Santiago de Compostela, E-15782\\ Santiago de Compostela, Spain, 
\email{xian.otero@usc.es}
}

%
%
\maketitle

\abstract{Phase transitions between two competing vacua of a given theory are quite common in physics. We discuss how to construct the space-time solutions that allow the description of phase transitions between different branches (or  asymptotics) of a given higher curvature gravity theory at finite temperature.}

\section{Introduction}
\label{intro}

Higher-curvature corrections to the Einstein-Hilbert (EH) action appear in any sensible theory of quantum gravity as next-to-leading orders in the effective action and some, {\it e.g.} the Lanczos-Gauss-Bonnet (LGB) action \cite{Lanczos}, also appear in realizations of string theory \cite{GBstrings1}. This quadratic combination is particularly important as any quadratic term can be brought to the LGB form, $\mathcal{R}^2=R_{\mu\nu\alpha\beta}R^{\mu\nu\alpha\beta}-R_{\mu\nu}R^{\mu\nu}+R^2$, via field redefinitions. 

Due to the non-linearity of the equations of motion, these theories generally admit more than one maximally symmetric solution, $R_{\mu \nu \alpha\beta}=\Lambda_i(g_{\mu\alpha}g_{\nu\beta}-g_{\mu\beta}g_{\nu\alpha})$; (A)dS vacua with effective cosmological constants $\Lambda_{i}$, whose values are determined by a polynomial equation \cite{BoulwareDeser}, 
\vskip-2mm
\begin{equation}
\Upsilon [\Lambda] \equiv \sum_{k=0}^{K}c_{k}\,\Lambda^{k} = c_{K}\prod_{i=1}^{K}\left( \Lambda -\Lambda _{i}\right) =0 ~.
\label{cc-algebraic}
\end{equation}
$K$ being the highest power of curvature (without derivatives) in the field equations. $c_0=1/L^2$ and $c_1=1$ give canonically normalized cosmological and EH terms, $c_{k\geq 2}$ are the LGB and higher order couplings (see \cite{JDEere} for details).

Any vacua is {\it a priori} suitable in order to define boundary conditions for the gravity theory we are interested in; {\it i.e.} we can define sectors of the theory as classes of solutions that asymptote to a given vacuum \cite{CE}. In that way, each branch has associated static solutions, representing either black holes or naked singularities,
\vskip-2mm
\begin{equation}
ds^{2}=-f(r)\,dt^{2}+\frac{dr^{2}}{g(r)}+r^{2}\ d\Omega_{d-2}^{2} ~, \qquad \qquad f,g \xrightarrow{r\rightarrow \infty} -\Lambda_i r^2 ~,
\label{bhansatz}
\end{equation}
and other solutions with the same asymptotics. The main motivation of the present work is that of studying transitions between different branches of solutions. This is important in order to investigate whether a new type of instability involving non-perturbative solutions occurs in the theory. This new kind of phase transitions have been recently investigated in the context of LGB \cite{Camanho2012} and Lovelock gravities \cite{comingsoon}. 

\section{Higher order free particle}

The existence of branch transitions in higher curvature gravity theories is a concrete expression of the multivaluedness problem of these theories. In general the  canonical momenta, $\pi_{ij}$, are not invertible functions of the velocities, $\dot{g}^{ij}$ \cite{Teitelboim1987}. An analogous situation may be illustrated by means of a simple one-dimensional example \cite{Henneaux1987b}. Consider a free particle lagrangian containing higher powers of velocities,
\vskip-1mm
\begin{equation}
L(\dot{x})=\frac{1}{2}\dot{x}^2-\frac13\dot{x}^3+\frac1{17}\dot{x}^4
\label{paction}
\end{equation}
In the hamiltonian formulation the equation of motion just implies the constancy of the conjugate momentum, $\frac{d}{dt}p=0$. However, being this multivalued (also the hamiltonian), the solution is not unique. Fixing boundary conditions $x(t_{1,2})=x_{1,2}$, an obvious solution would be constant speed 
$
\dot{x}=(x_2-x_1)/(t_2-t_1)\equiv v
$
but we may also have jumping solutions with constant momentum and the same mean velocity. 

\begin{figure}[h]
\sidecaption[r]
\includegraphics[scale=.67]{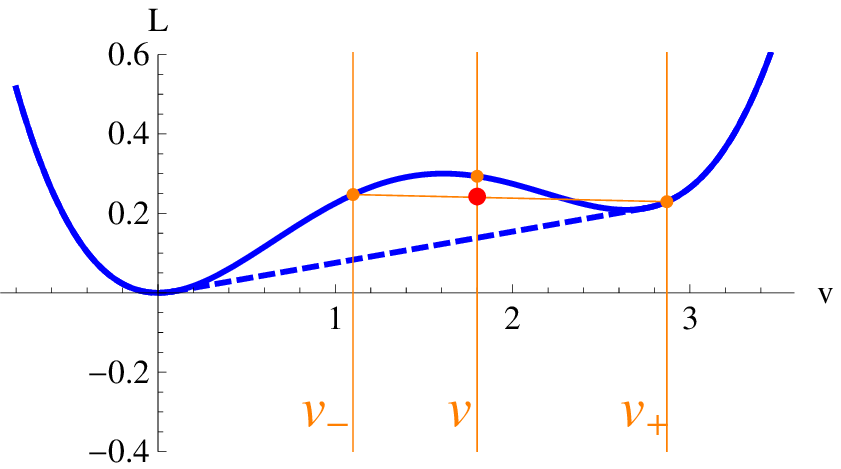}  \quad
\includegraphics[scale=.67]{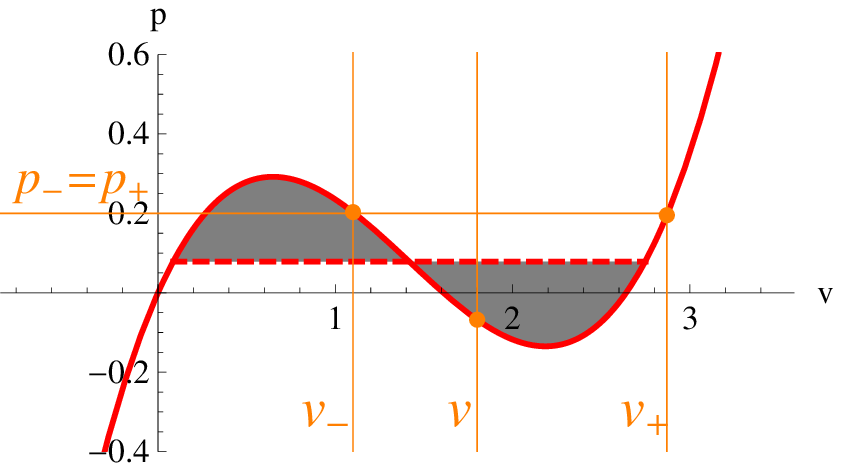} 
%
%
\caption{Lagrangian and momentum for the action (\ref{paction}). For the same mean velocity $v$, the action is lower for jumps between $v_\pm$ (big dot) than for constant speed, the minimum action corresponding to the value on the dashed line ({\it effective} Lagrangian).}
\label{fig:1}       
\end{figure}

In our example, for mean velocities  corresponding to multivalued momentum (see figure \ref{fig:1}) solutions are infinitely degenerate as the jumps may occur at any time and unboundedly in number as long as the mean velocity is the same. Nevertheless, this degeneracy is lifted once the value of the action is taken into account. The minimal action path is the naive one for mean velocities outside the range covered by the dashed line whereas in that interval it corresponds to arbitrary jumps between the  velocities of the two extrema. The {\it effective} Lagrangian (dashed line) is a convex function of the velocities and the effective momentum dependence corresponds to the analogous of the Maxwell construction from thermodynamics (see \cite{comingsoon} for a detailed explanation of this one-dimensional example).


\section{Generalized Hawking-Page transitions}

In the context of General Relativity in asymptotically AdS spacetimes, the Hawking-Page phase transition \cite{HawkingPage} is the realization that above certain temperature the dominant saddle in the gravitational partition function comes from a black hole, whereas for lower temperatures it corresponds to the thermal vacuum. The {\it classical} solution is the one with least Euclidean action among those with a smooth Euclidean section. 

When one deals with higher curvature gravity there is a crucial difference that has been overlooked in the literature. In addition to the usual continuous and differentiable metrics (\ref{bhansatz}), one may construct distributional metrics by gluing two solutions corresponding to different branches across a spherical shell or {\it bubble} \cite{wormholes,wormholes2}. The resulting solution will be continuous at the bubble --with discontinuous derivatives, even in absence of matter. The higher curvature terms can be thought of as a sort of matter source for the Einstein tensor. The existence of such {\it jump} metrics, as for the one-dimensional example, is due to the multivaluedness of momenta in the theory.

In the gravitational context, continuity of momenta is equivalent to the junction conditions that need to be imposed on the bubble. In the EH case, Israel junction conditions \cite{Israel1967}, being linear in velocities, also imply the continuity of derivatives of the metric. The generalization of these conditions for higher curvature gravity contain higher powers of velocities, thus allowing for more general situations.

Static bubble configurations, when they exist, have a smooth Euclidean continuation. It is then possible to calculate the value of the action and compare it to all other solutions with the same asymptotics and temperature. This analysis has been performed for the LGB action \cite{Camanho2012} for unstable boundary conditions \cite{BoulwareDeser}. The result suggests a possible resolution of the instability through bubble nucleation.

In the case of LGB gravity there are just two possible static spacetimes to be considered in the analysis for the chosen boundary conditions; the thermal vacuum and the static bubble metric, the usual spherically symmetric solution (\ref{bhansatz}) displaying a naked singularity. For low temperatures the thermal vacuum is the preferred solution whereas at high temperatures the bubble will form, as indicated by the change of sign on the relative free energy. The bubble pops out in an unstable  position and may expand reaching the boundary in a finite time thus changing the asymptotics and charges of the solution, from the initial to the inner ones.

Still, if the free energy is positive the system is metastable. It decays by nucleating bubbles with a probability given, in the semiclassical approximation, by $e^{-F/T}$. Therefore, after enough time, the system will always end up in the stable horizonful branch of solutions, the only one usually considered as relevant. This is then a natural mechanism that selects the general relativistic vacuum among all the possible ones, the stable branch being the endpoint of the initial instability.

\section{Discussion}

The phenomenon described here is quite general. It occurs also for general Lovelock gravities \cite{comingsoon} and presumably for more general classes of theories. In the generic case, however, the possible situations one may encounter are much more diverse. We may have for instance stable bubble configurations as opposed to the unstable ones discussed above or even bubbles that being unstable cannot reach the boundary of the spacetime.  Other generalizations may include transitions between positive and negative values of $\Lambda_i$ and even non-static bubble configurations. 

Another situation one may think of is that of having different gravity theories on different sides of the bubble. This has a straightforward physical interpretation if we consider the higher order terms as sourced by other fields that vary accross the bubble. For masses above $m^2>\|\Lambda_{\pm}\|$ a bubble made of these fields will be well approximated by a thin wall and we may integrate out the fields for the purpose of discussing the thermodynamics. If those fields have several possible vacuum expectation values leading to different theories we may construct interpolating solutions in essentially the same way discussed above. In this case the energy carried by the bubble can be interpreted as the energy of the fields we have integrated out. 

\begin{acknowledgement}
The author thanks A. Gomberoff for most interesting discussions, and the Front of pro-Galician Scientists for encouragement. He is supported by a spanish FPU fellowship. This work is supported in part by MICINN and FEDER (grant FPA2011-22594), by Xunta de Galicia (Conseller\'{\i}a de Educaci\'on and grant PGIDIT10PXIB206075PR), and by the Spanish Consolider-Ingenio 2010 Programme CPAN (CSD2007-00042).
\end{acknowledgement}



\begin{thebibliography}{99}

\bibitem{Lanczos}
C.~Lanczos, Annals Math.\  {\bf 39}, 842 (1938).

\bibitem{GBstrings1}
B.~Zwiebach, Phys.\ Lett.\ B {\bf 156}, 315 (1985).

\bibitem{BoulwareDeser}
D.~Boulware and S.~Deser, Phys.\ Rev.\ Lett.\  {\bf 55}, 2656 (1985).

\bibitem{JDEere}
J.~D.~Edelstein, these proceedings (and references therein).

\bibitem{CE}
X.~O.~Camanho and J.~D.~Edelstein, Class. Quantum Grav. {\bf 30}, 035009 (2013).

\bibitem{Camanho2012}
X.~O.~Camanho, J.~D.~Edelstein, G.~Giribet, A.~Gomberoff,
Phys.\ Rev.\ D {\bf 86}, 124048 (2012)

\bibitem{comingsoon}
X.~O.~Camanho {\it et al}., to appear.

\bibitem{Teitelboim1987}
C.~Teitelboim and J.~Zanelli, 
Class.\ Quantum Grav., {\bf 4}, 125 (1987).

\bibitem{Henneaux1987b}
M.~Henneaux, C.~Teitelboim, and J.~Zanelli, Phys.\ Rev.\ A, {\bf 36}, 4417 (1987).

\bibitem{HawkingPage}
S.~W.~Hawking and D.~N.~Page, Commun.\ Math.\ Phys.\  {\bf 87}, 577 (1983).

\bibitem{wormholes}
E. Gravanis and S. Willison, Phys. Rev. D {\bf 75}, 084025 (2007).

\bibitem{wormholes2}
C.~Garraffo, G.~Giribet, E.~Gravanis and S.~Willison, J.\ Math.\ Phys.\  {\bf 49}, 042502 (2008).

\bibitem{Israel1967}
W.~Israel, Nuovo Cimento B {\bf 48}, 463 (1967).
\end{thebibliography}



\end{document}